\documentclass{article}

\usepackage[final]{nips_2017}

\usepackage[utf8]{inputenc} % allow utf-8 input
\usepackage[T1]{fontenc}    % use 8-bit T1 fonts
\usepackage{hyperref}       % hyperlinks
\usepackage{url}            % simple URL typesetting
\usepackage{booktabs}       % professional-quality tables
\usepackage{amsfonts}       % blackboard math symbols
\usepackage{nicefrac}       % compact symbols for 1/2, etc.
\usepackage{microtype}      % microtypography
\usepackage{float}
\usepackage{graphicx}

\title{Generalized Seismic Phase Detection with Deep Learning}

\author{
  Zachary E. Ross\thanks{Correspondence to: \texttt{zross@gps.caltech.edu}} , Men-Andrin Meier, Egill Hauksson, Thomas H. Heaton \\
  Seismological Laboratory\\
  California Institute of Technology\\
  Pasadena, CA 91125 \\
}

\begin{document}
% \nipsfinalcopy is no longer used

\maketitle

\begin{abstract}

To optimally monitor earthquake-generating processes, seismologists have sought to lower detection sensitivities ever since instrumental seismic networks were started about a century ago. Recently, it has become possible to search continuous waveform archives for replicas of previously recorded events (template matching), which has led to at least an order of magnitude increase in the number of detected earthquakes and greatly sharpened our view of geological structures. Earthquake catalogs produced in this fashion, however, are heavily biased in that they are completely blind to events for which no templates are available, such as in previously quiet regions or for very large magnitude events. Here we show that with deep learning we can overcome such biases without sacrificing detection sensitivity. We trained a convolutional neural network (ConvNet) on the vast hand-labeled data archives of the Southern California Seismic Network to detect seismic body wave phases. We show that the ConvNet is extremely sensitive and robust in detecting phases, even when masked by high background noise, and when the ConvNet is applied to new data that is not represented in the training set (in particular, very large magnitude events). This generalized phase detection (GPD) framework will significantly improve earthquake monitoring and catalogs, which form the underlying basis for a wide range of basic and applied seismological research.
\end{abstract}

\section{Introduction}

Ideally, seismologists would be able to detect every single earthquake that occurs to optimally monitor seismogenic processes, such as fluid migration, elastic strain build-up and accelerating crustal deformation episodes. In practice, however, the ability to detect earthquakes is strongly limited by the fact that the vast majority of earthquakes are very small \cite{gutenberg_b._seismicity_1954}, and by the perpetual presence of a wide range of nuisance signals that are not caused by local earthquakes. 

Over the past decades, improvements in seismometer design, installation and network density have led to dramatic improvements in our capabilities for detecting small earthquakes. Methodological advances in earthquake detection and characterization continue to improve the resolution with which we can observe the seismically active parts of the crust. The magnitude of completeness—the magnitude above which all events have been detected by a network—for southern California now is \textasciitilde M1.8 in most areas \cite{hutton_earthquake_2010}.

Among the most successful earthquake detection algorithms developed to-date are those that exploit the similarity of earthquake waveforms between similarly located events. The most widely used version of these methods is template matching \cite{peng_migration_2009,skoumal_earthquakes_2015,shelly_fluid-faulting_2016,ross_aftershocks_2017}. By correlating the seismic waveforms of cataloged individual events against continuous waveform data, it is often possible to detect an order of magnitude more events than with routine methods \cite{allen_automatic_1982}. These similarity-based methods typically have additional stringent detection requirements, such as forcing the detected events to have the same move-out pattern as the template event across the network. While this makes the detections reliable, it also makes the similarity-based methods blind to what they have not seen before, such as events in previously inactive regions, or large magnitude earthquakes. As a consequence, catalogs created with such methods are inherently incomplete in that they contain only the subset of events that meet the rather specific detection criteria.

The task of recognizing seismic phases in a waveform time-series is very similar to that of recognizing objects in 2D images. A 3-component seismogram can be thought of as a 1D image, with the three components being equivalent to the three RGB color channels. The stunning recent advances in the field of computer vision therefore have important potential in seismological applications. Deep learning algorithms, which use hierarchically-organized non-linear mapping functions for tasks such as classification and regressions \cite{lecun_gradient-based_1998}, have been shown to be powerful tools for object recognition because of their ability to develop robust generalized representations of extremely large datasets \cite{he_deep_2015,krizhevsky_imagenet_2012}. They have become the state-of-the-art in various machine learning domains, including handwriting recognition, language processing and object categorization \cite{lecun_deep_2015}. Rather than using explicit template objects to search for in an image (e.g. template images of cats) these algorithms are able to detect the general characteristics that individual realizations of that object class share (e.g. the furry texture). With such non-explicit search they are able to reliably detect objects without ever having seen an exact or even similar object template.

In seismology, deep learning has recently been used to detect and roughly locate earthquakes in Oklahoma \cite{perol_convolutional_2018}, and to determine P-wave arrival times and first-motion polarities \cite{ross_z._e._p-wave_2018}. Here we develop a convolutional neural network that can detect and classify seismic body wave phases over a broad range of circumstances. We use the vast hand-labeled data archives of the Southern California Seismic Network. We show that this method for generalized phase detection (GPD) is capable of reliably detecting P- and S-waves of very small to very large earthquakes without the need for explicit waveform templates. The ability to generalize entire waveform and metadata archives enables the method to successfully operate on data that was not covered by the training data, including data from entirely different tectonic regimes, and of earthquakes much larger than the largest events in the training set. The developed procedure will allow us to compile fundamentally improved seismicity catalogs with detection sensitivities comparable to template matching catalogs, but without their inherent biases. Furthermore, it has the potential to make earthquake early warning systems more reliable.

\section{Methods and Results}
\label{headings}
\subsection{Neural Networks and Deep Learning}
In the most general sense, neural networks map a set of input values (e.g. a raster image or a time series) into a set of output values (e.g. the probability that the input image contains a cat, or that a seismogram contains a P-wave). The network architecture, which parameterizes the non-linear mapping function, has a large number of parameters (typically >1e6 for deep networks) that are empirically optimized with large amounts of training data. The optimization is performed such that the network predictions (i.e. the mapping output) are as accurate as possible across a large data set.

For standard fully-connected neural networks (FCNN), the mapping function consists of a large number of addition and multiplication operations, as well as simple non-linear activation functions. These operations are organized into discrete layers where the output of each layer is sequentially used as input to the next. The first layer of a FCNN acts on "features", which are attributes of the raw data, such as peak amplitudes and frequency measures. The choice of the features is usually somewhat arbitrary and difficult to optimize. Convolutional neural networks (ConvNets), on the other hand, typically combine a FCNN with a learnable and hierarchical feature extraction system (Fig. 1) that is trained to extract the relevant information directly from the raw data.

\begin{figure}[h]
  \centering
  \includegraphics[width=\textwidth]{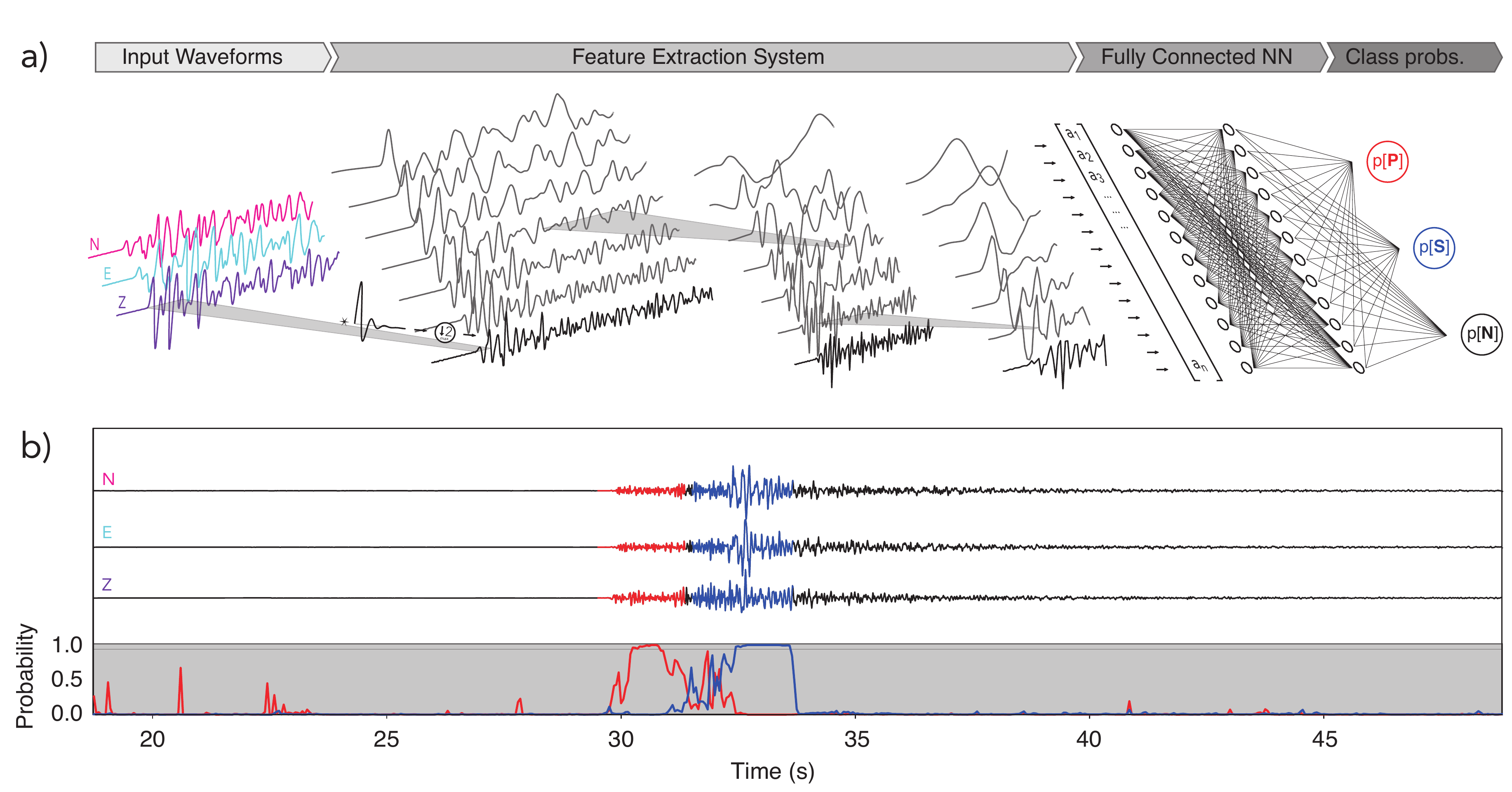}
  \caption{(a) Illustration of a convolutional network for generalized phase detection and (b) application example. By applying a series of filtering and decimation operations, features are automatically extracted and used to classify data windows. (b) Example usage of GPD to continuous data. Probability time series for P- (red) and S-waves (blue) are used as characteristic functions for phase detection. Detection windows with probability greater than 0.98 are colored by the respective phase type.}
\end{figure}

The feature extraction system itself consists of a series of layers in which the input data is sequentially i) convolved with a set of digital filters, ii) decimated, and iii) "activated". The digital filters parameterize the general object characteristics that the ConvNet will learn to recognize. Whenever this characteristic is present at some point in an input image, the convolution will yield a significant output value. The decimation, or "pooling", down-samples the image and causes the ConvNet to evaluate the image at different length scales. This enables the network to recognize the object regardless of its size, i.e. whether it covers the entire image, or whether it is just a small detail in the background. Finally, an activation function is used on the filtered and pooled input data that forces the output values to be positive-only, and that introduces a non-linearity in the mapping function. After several sequential convolution, pooling and activation steps, the resulting output of the last layer is concatenated to a vector and is used as input to a FCNN in order to perform regression or classification tasks. During the training process, the parameters of the entire network are learned directly from the data to maximize the network’s performance. This includes the coefficients of the digital filters and the coefficients of the linear functions of the FCNN’s neurons. 

Due to the convolutional nature of the filtering operations and the regular decimation, a ConvNet can detect a target object (e.g. a cat or a P-wave) irrespective of where it is located in an image or seismogram, or the size within the image. ConvNets are therefore ideal for seismic phase detection since the phases may occur anywhere in a seismogram, and can vary by orders of magnitudes in terms of duration and amplitude. The major limitation is that millions of labeled data samples are necessary to train the networks, which frequently have millions of free parameters.

\subsection{A ConvNet for Generalized Seismic Phase Detection}
We designed a convnet to scan through continuous seismic data, and for each 4 sec data window, classify the dominant category of the window as either a P-wave, S-wave, or noise (Figure 1b). The convnet was trained using millions of hand-labeled phases and their arrival times from analysts at the Southern California Seismic Network (see methods). The model architecture that we used consists of 6 layers (Table S1), which includes 4 convolution layers and 2 fully-connected layers. Rectified linear units were used as the activation function on each layer \cite{nair_rectified_2010}, and batch normalization was applied to each layer \cite{ioffe_batch_2015}. We explored many other possible models and found that overall, excellent performance was possible with both shallower and deeper models; thus, the method is generally insensitive to the specific architecture used. However, the network as described here performed the best of all those tested.

A total of 4.5 million 3-component seismic records were used for training and validation of the generalized phase detection framework. The data first were detrended and high-pass filtered above 2 Hz to remove microseismic noise, and all data were resampled at 100 Hz. Strong-motion records were integrated to velocity. These 3-component records consist of 1.5 million P-wave seismograms, 1.5 million S-wave seismograms, and 1.5 million noise windows, with each being exactly 4 s (400 samples) in duration. The magnitude range of the data was $-0.81 < M < 5.7$, while only records with epicentral distances less than 100 km were used. P- and S-wave windows were centered on the respective analyst pick, while noise windows were defined starting 5 s before each P-wave pick. These windows form the set of features used as input to the convnet, and examples of each class can be seen in Figure S2. The even distribution of records between each class ensures that the training process is not biased towards any one class. Then, each 3-component record was normalized by the absolute maximum amplitude observed on any of the three components. No other pre-processing was performed on the datasets.

To train the model, we first randomly split the seismograms into a training set (75\%) and validation set (25\%). Then, the model was trained using a cross-entropy loss function with the ADAM optimization algorithm \cite{kingma_adam:_2014}, in mini-batches of 480 records with 3 NVIDIA GTX 1060 graphical processing units. We programmed the training process to terminate when the validation loss had not decreased in more than 5 epochs (full iterations through the dataset), and the epoch with the lowest loss value on the validation set was selected as the best model (Figure S1). For this study, the 3rd epoch had the highest prediction accuracy on the validation set (99\%), with subsequent epochs resulting in less accurate predictions due to model overfitting. This forms the final model used throughout the paper.

We found that even though a high-pass filter at 2 Hz was used to train the model, that the method still worked over a broad range of frequency values above and below 2 Hz. For our subsequent testing with seismograms of the 2016 $M_w$ 7.0 Kumamoto, Japan earthquake, the method reliably detected P- and S-waves even when using acceleration traces (rather than velocity), and a high-pass filter at 0.1 Hz. Thus, it is not a strict requirement that the same filter used to train the data need be applied to new data for calculations.

\subsection{Classification Performance on the Validation Set}
First, we demonstrate the classification performance of the trained model on an independent validation set of 1.1 million seismograms split evenly between the three classes. For each 3-component seismogram, the model outputs three probabilities corresponding to the likelihood of each respective class (P, S or noise). A seismic phase detection is declared if the probability for either P or S is above a threshold level. If both predictions are below the threshold, the waveform is assigned to the noise class. Figure 2 shows the performance of the convnet, in terms of precision and recall, for a range of probability thresholds. While precision gives the fraction of correct declarations (true positives) relative to all declarations that have been made (true positives + false positives), recall gives the fraction of correct declarations relative to all declarations that should have been made (true positives + false negatives). 
\begin{figure}[h]
  \centering
  \includegraphics[width=\textwidth]{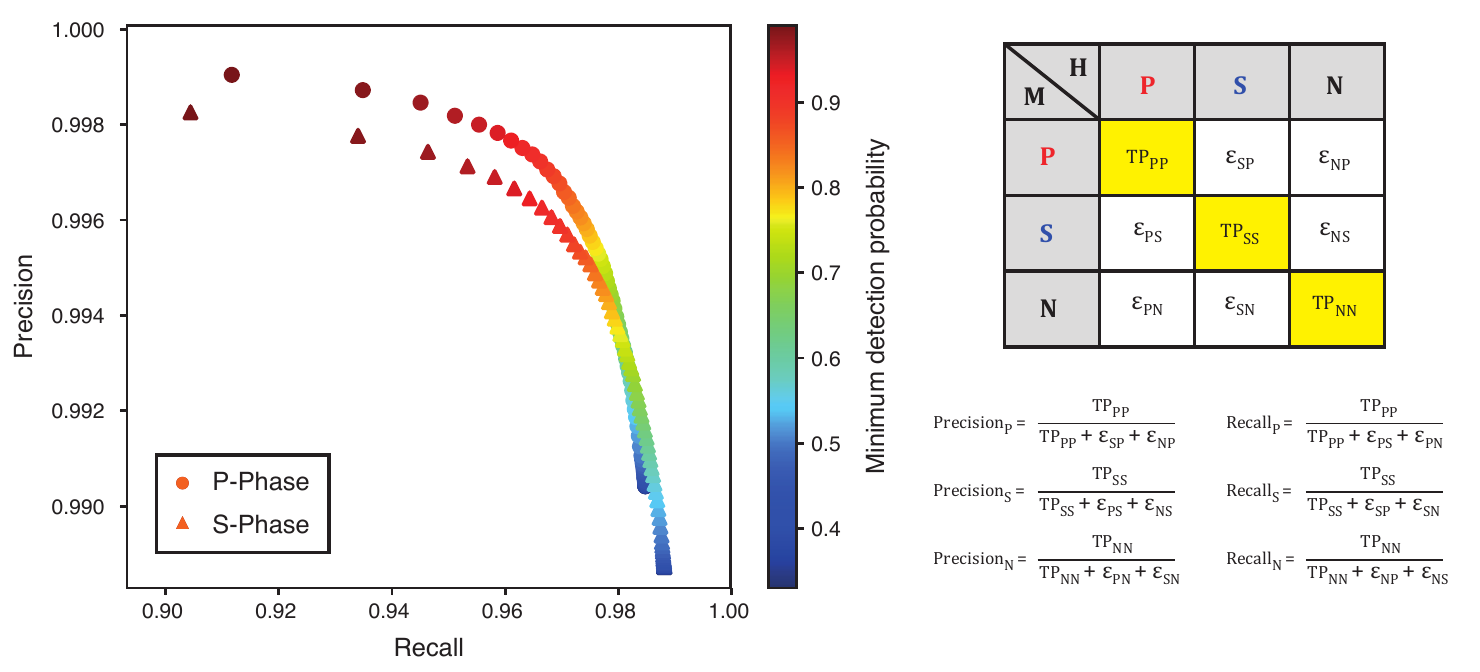}
  \caption{Precision-recall tradeoff curve for different declaration probability thresholds and confusion matrix with definitions. If P- or S-probabilities exceed the threshold probability (color) the waveform is assigned to the respective class. If neither probability exceeds the threshold the waveform is assigned to the noise class. For low probability thresholds more cases of false positives occur, while with higher thresholds more false negative cases occur. The confusion matrix shows the possible combinations of classifications by human analysts ('H') and the convnet algorithm ('A').}
\end{figure}

Almost irrespective of the detection threshold, the precision is very high (>99\%) for both phases, suggesting that that the network very rarely labels noise as seismic phases, or confuses P- and S-phases. Recall is somewhat lower, between 96\% and 99\% for most threshold choices, indicating that seismic phases are sometimes misclassified as noise. Remarkably, the performance is almost as good for S-waves as it is for P-waves.

These tradeoff curves illustrate that the probability threshold is a tunable parameter that can be exploited to serve a specific application. For earthquake catalog generation, for instance, one may wish to minimize false detections, while in an earthquake early warning (EEW) it may be more important to not have any missed detections. Furthermore, if one can further reduce false positives through other means (e.g. by requiring multiple stations to declare a detection), lower probability thresholds can be chosen in order to gain higher recall values (fewer false negatives). Figure S3 demonstrates the performance of the method as a function of SNR.

\subsection{Application to the 2016 Bombay Beach, California Swarm}
We next demonstrate the applicability of the GPD framework to detecting phases in a streaming mode during the 2016 Bombay Beach, California swarm \cite{hauksson_evolution_2017}. The trained network is used to classify 4 s windows of 3-component data (every 10 samples; 97.5\% overlap) over the first 24 h of the swarm. For each window, if the network predicts a P- or S-wave as the most likely category, and the forecast probability is greater than 0.98, the time point of the center of the window is assigned a detection with the respective phase label. When multiple consecutive values are above 0.98, the time point of the maximum value is taken as the detection time.

Due to the intensity of seismic activity during the swarm, we focus first on the initial 10 minute period following the onset. Figure 3 contains the 3-component velocity data at station BOM, which is located only 7 km away from the swarm itself. Here, P- and S-waves detected are colored red and blue, respectively. The forecast probability for these classes is shown as a function of time in the same respective colors. There are many earthquakes occurring within this short time window, and to more clearly illustrate the detection capabilities, we zoom in further on a segment that is just over two minutes long. It can be seen that during this short time window alone, 13 earthquakes are detected, and both phases are identified in 12 of them. The detections show higher P-wave amplitudes on the vertical component, and higher S-wave amplitudes on the horizontal components. S-wave detections are regularly preceded by P-wave detections, and S-P times are consistent with the largest events in the swarm, all of which suggest that the detections are real. The events are, on average, about 10 seconds apart; however, 6 of the events are significantly overlapping in time. We note that while the window duration used for phase detection here is 4 s long, the close proximity of station BOM to the swarm results in S-P times of about 1 s; thus, the method is able to robustly identify individual phases from very small segments of data, even when the phases overlap. An STA/LTA algorithm \cite{allen_automatic_1978,allen_automatic_1982} would trigger on few, if any, of these signals because the long-term average in the denominator does not reach small values during periods of elevated activity. Even a human analyst would likely have missed many of these events. The detection results are not unique to this 10 minute window; Figure 4 demonstrates nearly 1000 events were detected over the first 12 hours of the swarm alone, which is \textasciitilde 10 times as many as listed in the SCSN catalog; however it should also be mentioned that most of the SCSN events were included in the training data. Notably, the onset time of the swarm is measurable with extraordinary precision. Thus, GPD shows exceptional sensitivity to detecting P- and S-waves, without requiring specific templates to be correlated. We note that if the threshold of 0.98 is lowered, more events will be detected, but this will also result in additional false positives (Fig. 2).

\begin{figure}[h]
  \centering
  \includegraphics[width=\textwidth]{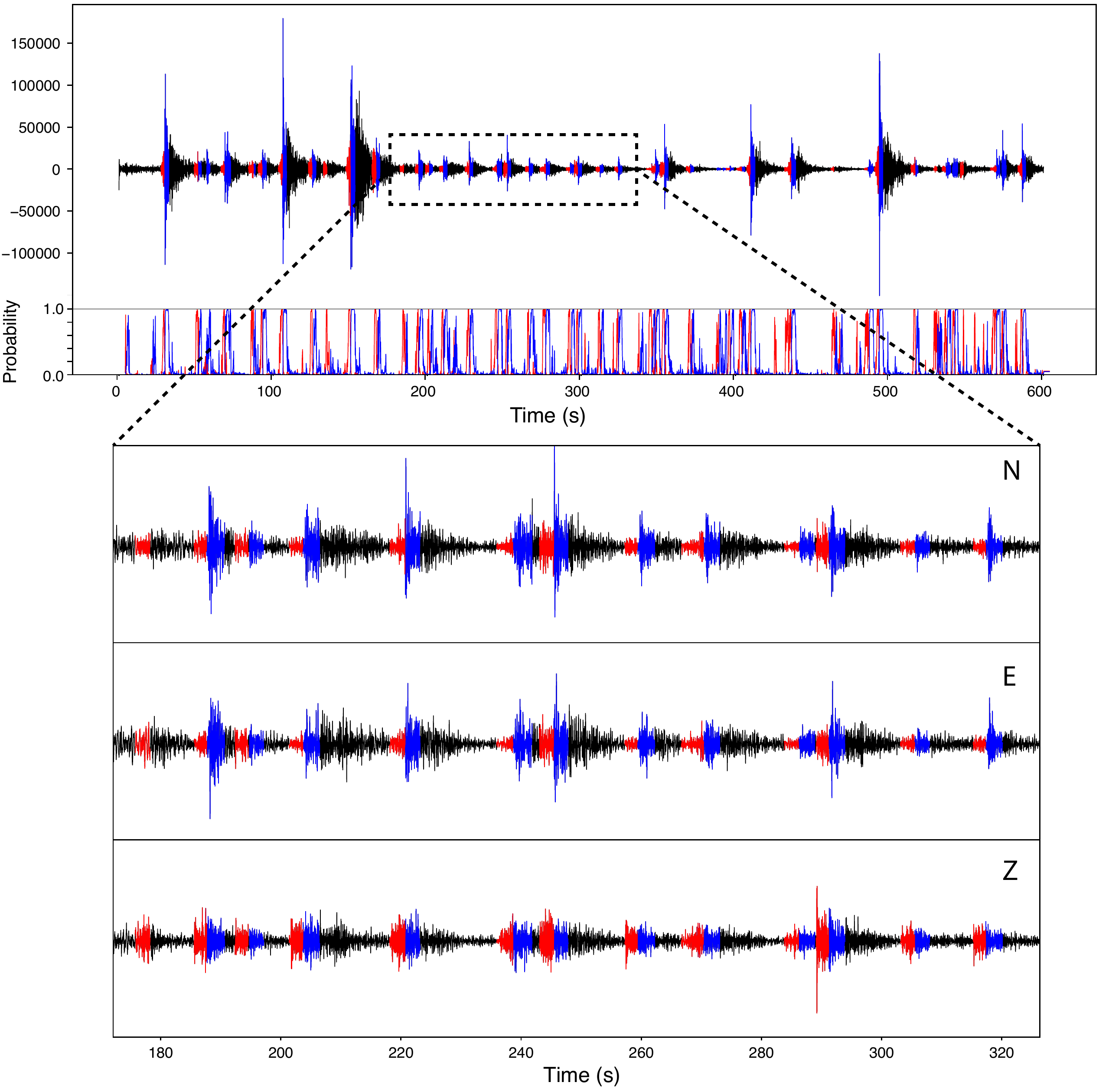}
  \caption{Application of GPD to the 2016 Bombay Beach swarm. P-wave (red) and S-wave (blue) detections are colored for all samples of any window in which a class probability exceeds 0.98. The probability time series shows numerous high-probability detections. Inset shows a close up of a time window \textasciitilde 140s long with 13 earthquakes detected. P-detections occur consistently before S-detections. While P-waves tend to have higher vertical amplitudes, S-waves are stronger on the horizontal components.}
\end{figure}

\begin{figure}[h]
  \centering
  \includegraphics[width=\textwidth]{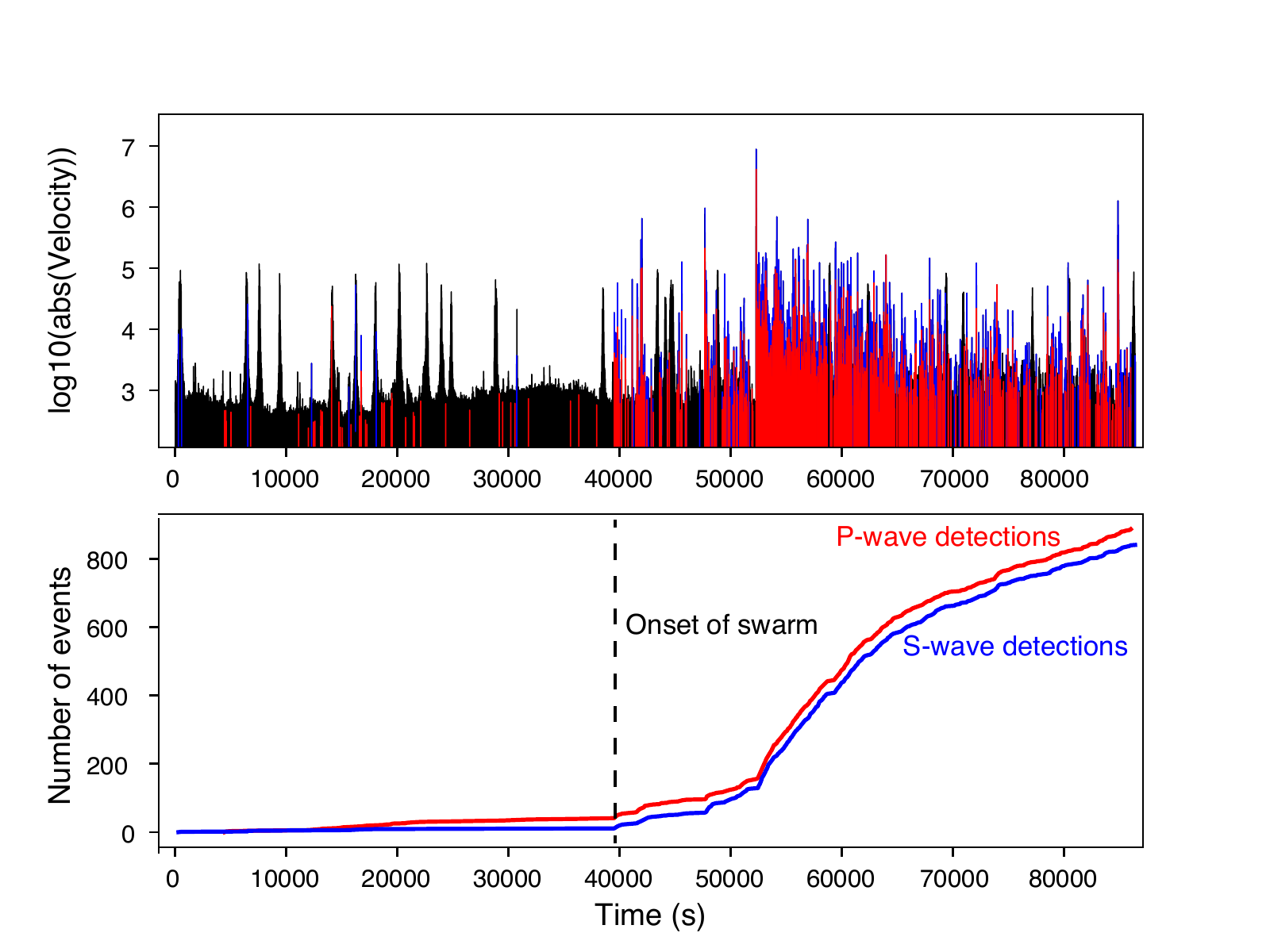}
  \caption{Example of GPD applied to the first 12 hours of the 2016 Bombay Beach sequence. The onset time of the swarm is sharply resolved. The total number of events detected is \textasciitilde 10 times as many as listed in the SCSN catalog. The swarm onset is distinctly visible, which often is not the case in routine catalogs, posing a common problem for seismic monitoring operators. The high amplitude signals before the swarm are nuisance signals from nearby vehicles that do not trigger the GPD algorithm.}
\end{figure}

\subsection{Application to the 2016 Mw 7.0 Kumamoto, Japan Earthquake}
Detection of large earthquakes is a critical task for seismic networks and earthquake early warning (EEW) systems, but methods that exploit the waveform similarity of previous events to detect new ones will be incapable of detecting these damaging events. We demonstrate the viability of GPD for detecting large earthquakes reliably by applying it to the 2016 Mw 7.0 Kumamoto earthquake \cite{asano_source_2016}. Figure 5 contains horizontal seismograms for all strong-motion stations within 100 km of the hypocenter. For every station, the P-wave of the mainshock is detected correctly, while for nearly every station, the S-wave is also detected. It is important to note that the training data were exclusively recorded in southern California, but here the model is detecting phases in Japan on completely different instruments. These results are even more remarkable given that the network was only trained on earthquakes with M < 5, and thus the model has not seen a large magnitude earthquake during the training process. These results are possible because within the short time windows used for prediction, the earthquake has not grown large enough in size yet to appear significantly different than a smaller event \cite{meier_evidence_2016}. Therefore, GPD identifies coherent arrivals, rather than entire waveforms, which enables detection of even large events.

\begin{figure}[h]
  \centering
  \includegraphics[width=\textwidth]{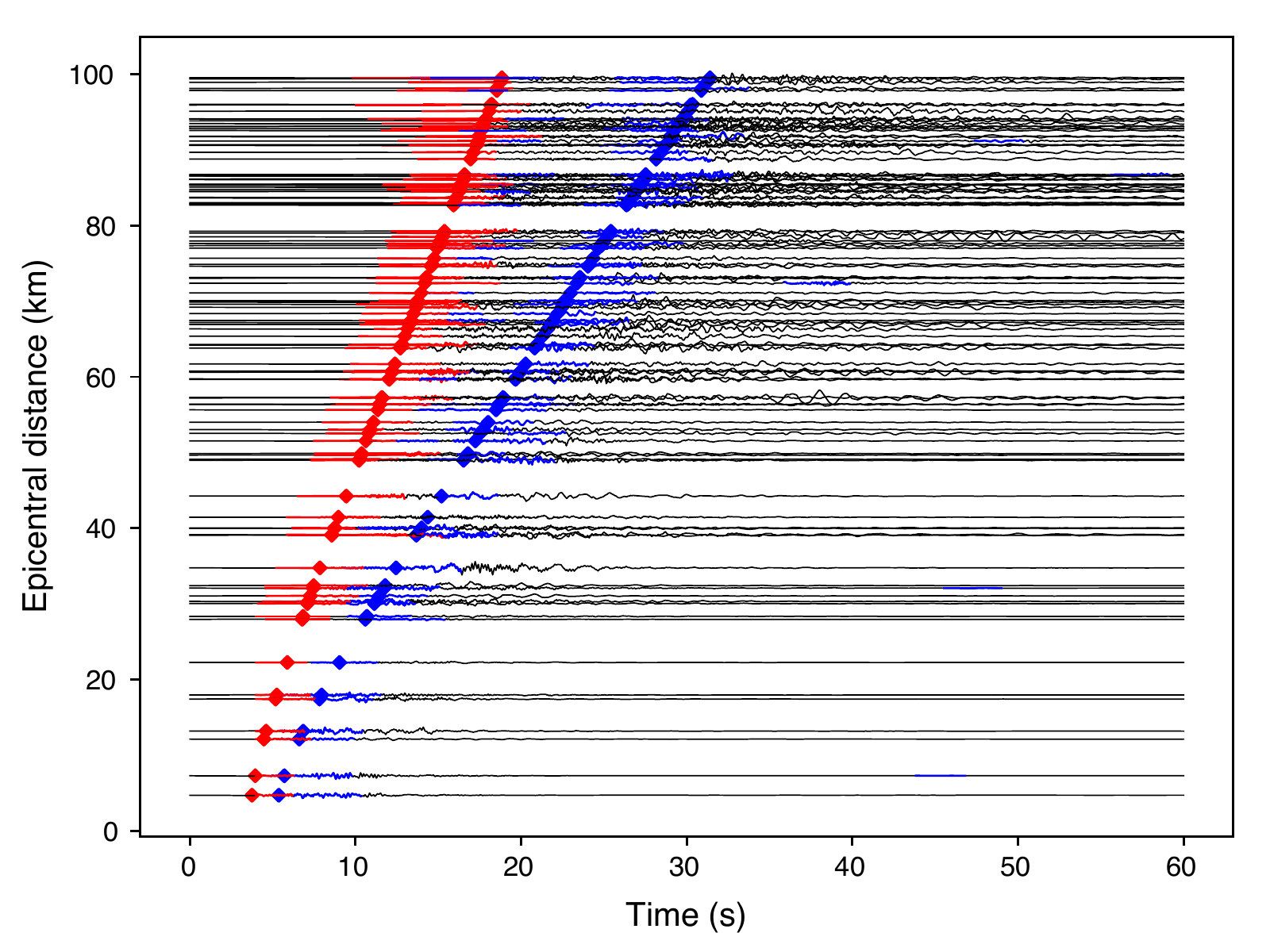}
  \caption{Application of GPD to the 2016 Kumamoto earthquake. P-waves (red) and S-waves (blue) are detected at nearly all stations over the epicentral distance range 3-100 km. Diamonds indicate theoretical arrival times. This demonstrates both the viability of GPD to detect large magnitude earthquakes, but also to work in regions and magnitude ranges that are not included in the training data.}
\end{figure}

\section{Discussion}
Detection of seismic phases is the first step in producing a seismicity catalog, and subsequent measurements derived from these phases form the basis for nearly all forms of seismological analyses. We showed that by using deep learning to develop generalized representations of millions of P-waves, S-waves, and noise records, it is possible to significantly improve this initial phase detection process. The method as presented here has detection sensitivity that rivals template matching but is not associated with any of the biases inherent to the similarity-based detection methods. Although it currently is not possible to understand how the convnet operates in significant detail, it is likely that the network learns to recognize attributes of the 3D particle motion both before and after the phase arrival, as well as some frequency-based information. The network seems to be primarily detecting onsets of phases, which is likely because the data windows were centered on the phase arrival time; this is probably the reason that the method can work on large magnitude earthquakes, since the onsets look similar to those events. If multiple phases are present within a data window, it tends to assign class labels based on whatever signal type is closest to the center of the window (e.g. Fig. 3). Rather than trying to formalize this operation into a parametric set of rules, the network learns the best way to discriminate between the three signal classes based on the data directly.

While the model as trained demonstrates outstanding portability, for some applications, the model would likely have to either be retrained using larger distance recordings, or adapted to incorporate additional signal classes. For the latter, since the noise windows were taken randomly from pre-event data, only a small fraction will include high amplitude transient signals like vehicles, calibration pulses, or noise spikes. Some of these signal types could therefore lead to occasional spurious classifications as P- or S-waves since they are poorly represented in the training data. GPD provides a natural framework for including these signal types as additional classes once a library of them is produced. This will help to both improve the P- and S-wave detection sensitivity while also improving the classification accuracy, since the decision boundaries between each signal type will become sharper. For regional distances and beyond, the presence of non-direct phases can potentially make detections more complicated and it may be desirable to eventually include these phase types explicitly. The model presented here is specifically trained to detect direct P- and S- phases at local distances.

An important potential application of GPD is in EEW, to improve discrimination between genuine earthquake signals and other spurious signals that could lead to false detections. Many EEW systems currently trigger based on large STA/LTA signals, which serve as an anchor for the real-time monitoring process \cite{cochran_earthquake_2018}. GPD can be applied in a streaming mode for each packet received in an analogous manner to STA/LTA. Every time a P-wave is detected, the system could trigger other event characterization schemes (e.g. magnitude estimation \cite{kanamori_real-time_2005}), and ultimately issue an alert if necessary. By looking for signals that are reminiscent of P-waves, rather than arbitrarily polarized large amplitude signals, GPD can provide a far more stable base to EEW systems. Furthermore, the additional information of what phase-type a detection corresponds to (P vs. S) is valuable information in an EEW context, since it can be used for excluding S-wave observations from magnitude-estimation schemes that have been optimized only for P-waves \cite{bose_real-time_2009} and to assess if an ongoing rupture is still growing \cite{kodera_real-time_2018}.

The next big question is how the GPD results can be optimally harnessed to generate substantially improved earthquake catalogs. The impressive classification performance and the outstanding sensitivity of the method as applied to individual stations suggests that once information from multiple stations is combined, event detection via phase association should be robust. Furthermore, the fact that GPD provides reliable phase labels (P vs. S) can be of great value for unscrambling the seismic signals from simultaneous and potentially overlapping events, such as during swarms.

In summary, the GPD framework may sharpen our ability to characterize how earthquakes evolve in space and time over all scales. It may help us to better characterize and understand seismogenic processes, from tiny micro-earthquakes whose signals barely exceed the ambient noise level, all the way to destructive large magnitude event sequences. 

\section{Data and Resources}
In this study, we used seismograms for 273,882 earthquakes $(-0.81 < M < 5.7)$ recorded by the Southern California Seismic Network (SCSN) at 692 broadband and short-period 3-component stations from 2000-2017 \cite{southern_california_earthquake_data_center_southern_2013}. The waveform data (last accessed January 2018) were associated with 1.5 million P-wave picks and 1.5 million S-wave picks that were manually determined by SCSN analysts. We also used 24 h continuous data at station CI BOM on 2016-09-26, the first day of the 2016 Bombay Beach, California swarm. Seismicity during this sequence was taken from the SCSN regional catalog. Finally, we used KiK-net and K-net accelerograms of the 2016 Mw 7.0 Kumamoto earthquake at stations within 100 km distance of the hypocenter (last accessed January 2018). The trained model and an accompanying dataset will be available through the Southern California Earthquake Data Center (scedc.caltech.edu).

\section{Acknowledgements}
This research was supported by grants from the Gordon and Betty Moore Foundation, the Swiss National Science Foundation and the NSF Geoinformatics program. We have used waveforms and meta-data data from Japanese networks and the Caltech/USGS Southern California Seismic Network (SCSN); doi: 10.7914/SN/CI; stored at the Southern California Earthquake Data Center. doi:10.7909/C3WD3xH1. We used Tensorflow to train the convolutional network.

\bibliographystyle{abbrv}
\bibliography{gpd_arxiv}

\begin{thebibliography}{10}

\bibitem{allen_automatic_1982}
R.~Allen.
\newblock Automatic phase pickers: {Their} present use and future prospects.
\newblock {\em Bulletin of the Seismological Society of America},
  72(6B):S225--S242, Dec. 1982.

\bibitem{allen_automatic_1978}
R.~V. Allen.
\newblock Automatic earthquake recognition and timing from single traces.
\newblock {\em Bulletin of the Seismological Society of America},
  68(5):1521--1532, Oct. 1978.

\bibitem{asano_source_2016}
K.~Asano and T.~Iwata.
\newblock Source rupture processes of the foreshock and mainshock in the 2016
  {Kumamoto} earthquake sequence estimated from the kinematic waveform
  inversion of strong motion data.
\newblock {\em Earth, Planets and Space}, 68(1):147, Dec. 2016.

\bibitem{bose_real-time_2009}
M.~Böse, E.~Hauksson, K.~Solanki, H.~Kanamori, and T.~H. Heaton.
\newblock Real-time testing of the on-site warning algorithm in southern
  {California} and its performance during the {July} 29 2008 {Mw}5.4 {Chino}
  {Hills} earthquake.
\newblock {\em Geophysical Research Letters}, 36(5):1--5, Mar. 2009.

\bibitem{cochran_earthquake_2018}
E.~S. Cochran, M.~D. Kohler, D.~D. Given, S.~Guiwits, J.~Andrews, M.-A. Meier,
  M.~Ahmad, I.~Henson, R.~Hartog, and D.~Smith.
\newblock Earthquake {Early} {Warning} {ShakeAlert} {System}: {Testing} and
  {Certification} {Platform}.
\newblock {\em Seismological Research Letters}, 89(1):108--117, Jan. 2018.

\bibitem{gutenberg_b._seismicity_1954}
B.~Gutenberg and C.~F. Richter.
\newblock {\em Seismicity of the {Earth}}.
\newblock Hafner Publishing Company, New York, 1954.

\bibitem{hauksson_evolution_2017}
E.~Hauksson, M.~Meier, Z.~E. Ross, and L.~M. Jones.
\newblock Evolution of seismicity near the southernmost terminus of the {San}
  {Andreas} {Fault}: {Implications} of recent earthquake clusters for
  earthquake risk in southern {California}.
\newblock {\em Geophysical Research Letters}, 44(3):1293--1301, 2017.

\bibitem{he_deep_2015}
K.~He, X.~Zhang, S.~Ren, and J.~Sun.
\newblock Deep {Residual} {Learning} for {Image} {Recognition}.
\newblock {\em arXiv:1512.03385 [cs]}, Dec. 2015.
\newblock arXiv: 1512.03385.

\bibitem{hutton_earthquake_2010}
K.~Hutton, J.~Woessner, and E.~Hauksson.
\newblock Earthquake {Monitoring} in {Southern} {California} for
  {Seventy}-{Seven} {Years} (1932–2008).
\newblock {\em Bulletin of the Seismological Society of America},
  100(2):423--446, Apr. 2010.

\bibitem{ioffe_batch_2015}
S.~Ioffe and C.~Szegedy.
\newblock Batch {Normalization}: {Accelerating} {Deep} {Network} {Training} by
  {Reducing} {Internal} {Covariate} {Shift}.
\newblock In {\em {PMLR}}, pages 448--456, June 2015.

\bibitem{kanamori_real-time_2005}
H.~Kanamori.
\newblock Real-{Time} {Seismology} and {Earthquake} {Damage} {Mitigation}.
\newblock {\em Annual Review of Earth and Planetary Sciences}, 33(1):195--214,
  2005.

\bibitem{kingma_adam:_2014}
D.~P. Kingma and J.~Ba.
\newblock Adam: {A} {Method} for {Stochastic} {Optimization}.
\newblock {\em arXiv:1412.6980 [cs]}, Dec. 2014.
\newblock arXiv: 1412.6980.

\bibitem{kodera_real-time_2018}
Y.~Kodera.
\newblock Real-{Time} {Detection} of {Rupture} {Development}: {Earthquake}
  {Early} {Warning} {Using} {P} {Waves} {From} {Growing} {Ruptures}.
\newblock {\em Geophysical Research Letters}, 45(1):2017GL076118, Jan. 2018.

\bibitem{krizhevsky_imagenet_2012}
A.~Krizhevsky, I.~Sutskever, and G.~E. Hinton.
\newblock {ImageNet} {Classification} with {Deep} {Convolutional} {Neural}
  {Networks}.
\newblock In F.~Pereira, C.~J.~C. Burges, L.~Bottou, and K.~Q. Weinberger,
  editors, {\em Advances in {Neural} {Information} {Processing} {Systems} 25},
  pages 1097--1105. Curran Associates, Inc., 2012.

\bibitem{lecun_deep_2015}
Y.~LeCun, Y.~Bengio, and G.~Hinton.
\newblock Deep learning.
\newblock {\em Nature}, 521(7553):436--444, May 2015.

\bibitem{lecun_gradient-based_1998}
Y.~LeCun, L.~Bottou, Y.~Bengio, and P.~Haffner.
\newblock Gradient-based learning applied to document recognition.
\newblock {\em Proceedings of the IEEE}, 86(11):2278--2324, Nov. 1998.

\bibitem{meier_evidence_2016}
M.-A. Meier, T.~Heaton, and J.~Clinton.
\newblock Evidence for universal earthquake rupture initiation behavior.
\newblock {\em Geophysical Research Letters}, 43(15):7991--7996, Aug. 2016.

\bibitem{nair_rectified_2010}
V.~Nair and G.~E. Hinton.
\newblock Rectified linear units improve restricted boltzmann machines.
\newblock In {\em Proceedings of the 27th international conference on machine
  learning ({ICML}-10)}, pages 807--814, 2010.

\bibitem{peng_migration_2009}
Z.~Peng and P.~Zhao.
\newblock Migration of early aftershocks following the 2004 {Parkfield}
  earthquake.
\newblock {\em Nature Geoscience}, 2(12):877--881, Dec. 2009.

\bibitem{perol_convolutional_2018}
T.~Perol, M.~Gharbi, and M.~Denolle.
\newblock Convolutional neural network for earthquake detection and location.
\newblock {\em Science Advances}, 4(2):e1700578, Feb. 2018.

\bibitem{ross_z._e._p-wave_2018}
Z.~E. Ross, M.-A. Meier, and E.~Hauksson.
\newblock P-wave arrival picking and first-motion polarity determination with
  deep learning.
\newblock {\em J. Geophys. Res.-Solid Earth, submitted}, 2018.
\newblock arXiv:1804.08804 [physics.geo-ph].

\bibitem{ross_aftershocks_2017}
Z.~E. Ross, C.~Rollins, E.~S. Cochran, E.~Hauksson, J.-P. Avouac, and
  Y.~Ben-Zion.
\newblock Aftershocks driven by afterslip and fluid pressure sweeping through a
  fault-fracture mesh.
\newblock {\em Geophysical Research Letters}, page 2017GL074634, Jan. 2017.

\bibitem{shelly_fluid-faulting_2016}
D.~R. Shelly, W.~L. Ellsworth, and D.~P. Hill.
\newblock Fluid-faulting evolution in high definition: {Connecting} fault
  structure and frequency-magnitude variations during the 2014 {Long} {Valley}
  {Caldera}, {California}, earthquake swarm.
\newblock {\em Journal of Geophysical Research-Solid Earth}, 121(3):1776--1795,
  Mar. 2016.

\bibitem{skoumal_earthquakes_2015}
R.~J. Skoumal, M.~R. Brudzinski, and B.~S. Currie.
\newblock Earthquakes {Induced} by {Hydraulic} {Fracturing} in {Poland}
  {Township}, {OhioEarthquakes} {Induced} by {Hydraulic} {Fracturing} in
  {Poland} {Township}, {Ohio}.
\newblock {\em Bulletin of the Seismological Society of America},
  105(1):189--197, Feb. 2015.

\bibitem{southern_california_earthquake_data_center_southern_2013}
{Southern California Earthquake Data Center}.
\newblock Scsn.
\newblock {\em California Institute of Technology, Dataset}, 2013.

\end{thebibliography}

\end{document}